# Measurement and theory of gravitational coupling between resonating beams

Tobias Brack[1*], Bernhard Zybach[1*], Fadoua Balabdaoui[2], Stephan Kaufmann[1], Francesco Palmegiano[1], Jean-Claude Tomasina[1], Stefan Blunier[1], Donat Scheiwiller[1], Jonas Fankhauser[1] & Jürg Dual[1 †]

[1] *Institute of Mechanical Systems, Department of Mechanical and Process Engineering, ETH Zurich, Tannenstrasse 3, 8092 Zürich*

[2] *Seminar for Statistics, Department of Mathematics, ETH Zurich, Rämistrasse 101, 8092 Zurich*

## Abstract

Recent spectacular results of gravitational waves obtained by the LIGO system[1], with frequencies in the 100 Hz regime, make corresponding laboratory experiments with full control over cause and effect of great importance. Dynamic measurements of gravitation in the laboratory have to date been scarce, due to difficulties in assessing non-gravitational crosstalk and the intrinsically weak nature of gravitational forces[2]. In fact, fully controlled quantitative experiments have so far been limited to frequencies in the mHz regime[3–6]. New experiments in gravity might also yield new physics, thereby opening avenues towards a theory that explains all of physics within one coherent framework.

Here we introduce a new, fully-characterized experiment at three orders of magnitude higher frequencies. It allows experimenters to quantitatively determine the dynamic gravitational interaction between two parallel beams vibrating at 42 Hz in bending motion.

The large amplitude vibration of the transmitter beam produces gravitationally-induced motion with amplitudes up to $10^{-11}$ m of the resonant detector beam. The reliable measurement with sub-pm displacement resolution is made possible by a set-up which combines acoustical, mechanical and electrical isolation, a temperature-stable environment, heterodyne laser interferometry and lock-in detection. The interaction is quantitatively modelled based on Newton's theory. Our initial results agree with the theory to within about three percent in amplitude. Based on a power balance analysis, we determined the near-field gravitational energy flow from the transmitter to the detector to be 2.5 $10^{-20}$ J/s, and to decay with distance as $d^{-4}$. We expect our experiment to make significant progress in directions where current experimental evidence for dynamic gravitation is limited, such as the dynamic determination of *G*, inverse-square law, and gravitational shielding[7].

---

[*] These authors contributed equally to this work
[†] Corresponding author: dual@imes.mavt.ethz.ch

# Main

Gravitation is the weakest of the four fundamental forces of physics. It has been modelled first by Newton[8], then extended with Einstein's theory of general relativity[9]. Experimentally, much progress has been made since Cavendish[2–5,10]. Nevertheless, many effects central to gravity have only been measured with a moderate accuracy due to the weakness of the gravitational force. Therefore, $G = 6.6743 \cdot 10^{-11}$ m$^3$kg$^{-1}$s$^{-2}$ is the fundamental physical constant known to the smallest precision with a relative uncertainty of 22 ppm[11]. Besides the determination of $G$, recent work has also focused on the investigation of the inverse square law[6,12,13] and the measurement of gravitational waves[14].

Measurements in the laboratory have the advantage of having full control over cause and effect without having to wait for something to happen in outer space[15]. Most experiments in this regard have been static or in the mHz regime[3,5,6]. Although some sources of uncertainty have been resolved, many complications remain (1/f noise, unknown static mass distributions, etc.) and uncertainties are still relatively large[3,4,16]. Also, there are still some yet unexplained differences in recent measurements that are assumed to be systematic errors. Hence, "The only way to give confidence is to measure the same constant using a number of different methods"[17].

Although the idea is fairly old[18,19], dynamic, resonance-based experiments using time-harmonic gravitational fields with frequencies > 0.1 Hz are very rare and start with Sinsky/Weber who observed an increase in vibrational noise of 20% due to gravitational interaction[20]. Hirakama et al. used a rotating quadrupole source to dynamically investigate the inverse square law (ISL)[21]. Astone et al. also used a rotating body to excite the gravitational wave antenna "Explorer" at CERN[22]. More recently, the use of micromechanical oscillators has been proposed in order to study the ISL at small distances[23] or to investigate the theory of quantum gravity using very small source masses[24,25].

None of these experiments, however, showed quantitative agreement between theory and experiment for all gravitational quantities at the same time and numerous engineering aspects have been mentioned that still need to be solved[24].

Around 1997, William Walker in our group worked for his Ph.D. thesis on a new dynamic setup consisting of two parallel beams both vibrating in the vicinity of the first resonant bending mode at a detector resonance frequency of about 40 Hz[26]. The brass transmitter beam was vibrating with a large amplitude of about 0.01 m in air, generating a dynamic gravitational force field. The tuned detector beam was set into vibration by this force field. The cylindrical detector beam made of quartz glass was placed at a distance of about 0.05 m in a vacuum chamber and had a quality factor of about 210000. This resulted in a gravitationally induced vibration amplitude of $10^{-9}$ m of the detector beam. However, because of the high time constant of the detector beam of ca. 90 min, difficulties in temperature stabilization made it impossible to make precise measurements. Many additional limitations in electronics, optics, and disturbing influences (mechanical and acoustic crosstalk, electromagnetic and magnetic effects, additional modes of vibration etc.) gave the impression that further work was not promising, and the project was abandoned.

In the meantime, many advances in signal processing, laser interferometry, materials technology, vibration isolation and control theory have been made. The setup was therefore revisited. Many improvements were implemented, and the setup's limitations were tested. These measures and the resulting outcome are presented here.

First the Newtonian theory is applied to the gravitational interaction between two beams to yield a vibrational amplitude. Then a working setup is described, and the measurement procedure is presented. In the results section detector amplitudes are shown for various distances and compared to the theory. Based on the theory, a first estimation of the dynamic gravitational constant $G$ is made based on 18 individual measurements, the inverse square law is tested, and the gravitational energy flow computed.

# Theory

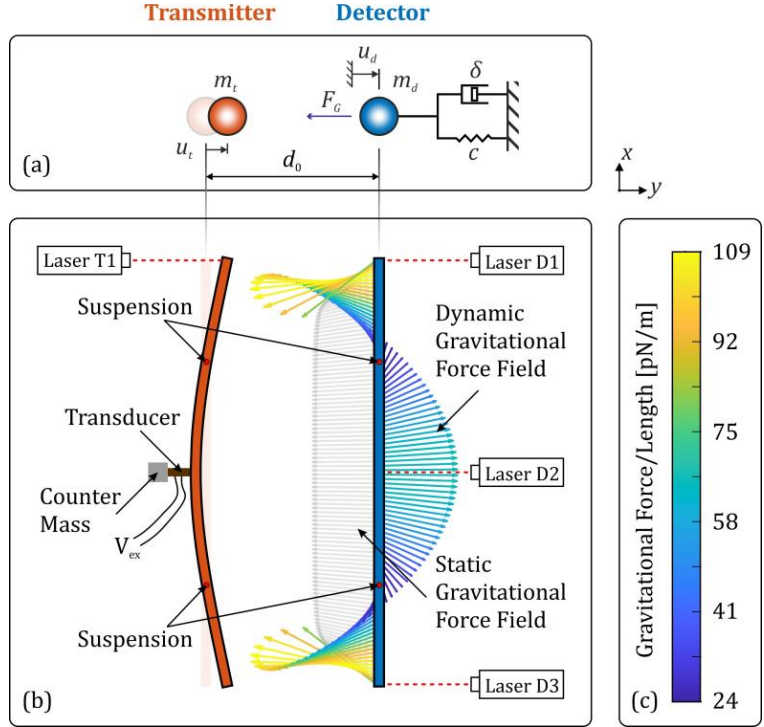

**Figure 1: Illustration of the measurement principle. a**, Simplified single-degree-of-freedom model with transmitter point mass $m_t$ and detector oscillator (point mass $m_d$, spring constant $c$, damping constant $\delta$): The oscillating transmitter mass creates a dynamic gravitational force $F_G$ acting on the detector mass thus generating an oscillation. When the frequency of excitation matches the resonance frequency of the oscillator, the detector will start to resonate and develop a measurable vibration amplitude. **b**, Sketch of the measurement setup using two bending beams: a piezo-transducer driven transmitter beam (orange) and detector beam (blue), both suspended in the nodal points of the first bending mode. The vibrating transmitter creates a gravitational force field on the detector, mainly composed of a static portion (grey colour, qualitatively) and a dynamic portion at the frequency of excitation (coloured). The vibration of the detector is measured by three laser vibrometers. **c**, Colourbar, illustrating the gravitational force density of the dynamic force field in the $xy$-plane in pN/m for the setup presented in this paper ($d_0$ = 59 mm).

Figure 1a illustrates two point masses, a transmitter mass $m_t$ and a detector mass $m_d$, separated by an undisturbed distance $d_0$, whereas the detector mass is part of a damped spring mass oscillator, characterized by its mass, the spring constant $c$ and damping constant $\delta$. When the transmitter mass moves with a time harmonic displacement $u_t(t) = u_t' e^{i\omega t}$, Newton's law of gravitation yields a dynamic gravitational force acting on both the transmitter and detector mass[8]. Due to the varying distance, the force is periodically changing around the static value $F_{G,0} = G m_d m_t d_0^{-2}$, thus causing the detector to vibrate.

When the frequency of excitation $\omega$ matches the resonance frequency $\omega_0 = (c/m_d)^{1/2}$ of the oscillator, the effect of the excitation is maximally amplified, resulting in a displacement of the detector mass that can be approximated by

$$u_d(t) \approx i \cdot \frac{2 G m_t Q_d}{d_0^3 \omega_0^2} u_t' e^{i\omega_0 t} \tag{1}$$

using a Taylor expansion under the assumption that $u_t' \ll d_0$ (cf. Methods). $Q_d \approx (c m_d)^{1/2} \delta^{-1}$ denotes the quality factor of the oscillator, assuming small damping. Equation (1) indicates that dynamic gravitational interaction can be observed if suitable parameters are chosen. Note that the detector displacement is not directly dependent on its mass. Consequently, the dynamic acceleration amplitude will be larger by a factor of $2 Q_d u_t'/d_0$ compared to the acceleration due to the gravitational force in a static setup.

The working principle as illustrated with point masses in Fig. 1a can be transformed to continuous bodies, such as vibrating bending beams, cf. Fig. 1b. In this case, both the transmitter and detector beam are resonating bending beams, hence both oscillators must have similar resonance frequencies to achieve sufficient amplitudes. For the setup and measurements described in this article we focus on the first bending mode, as illustrated in Fig. 1b. To theoretically describe the interaction between the two beams, Newton's law of gravitation is applied in terms of continuum mechanics with a gravitational force field acting on the detector beam (cf. Methods). Like the point-mass model, said force field is composed of a static part and a dynamic part at the frequency of excitation as well as at higher harmonics.

Applying the resulting force distribution as excitation force to the well-known equation of motion of a free-free Euler-Bernoulli beam yields the velocity amplitude of the detector beam at resonance that can be formulated analogously to the simple point mass model via

$$\frac{v_{d,0}}{v_{t,0}} \approx \mathrm{i} \cdot \frac{G m_t Q_d}{\omega_0^2} \Gamma(d_0, \text{parameters}) \qquad (2)$$

where $v_{d,0}$ and $v_{t,0}$ describe the (complex) velocity amplitudes of the detector and transmitter bending resonance motion, respectively. The function $\Gamma$ is dependent on the distance $d_0$, the beams dimensions, additional moving masses and their relative positions and can be calculated using an analytical or finite element 3D model of the setup (cf. Methods).

## Experiment

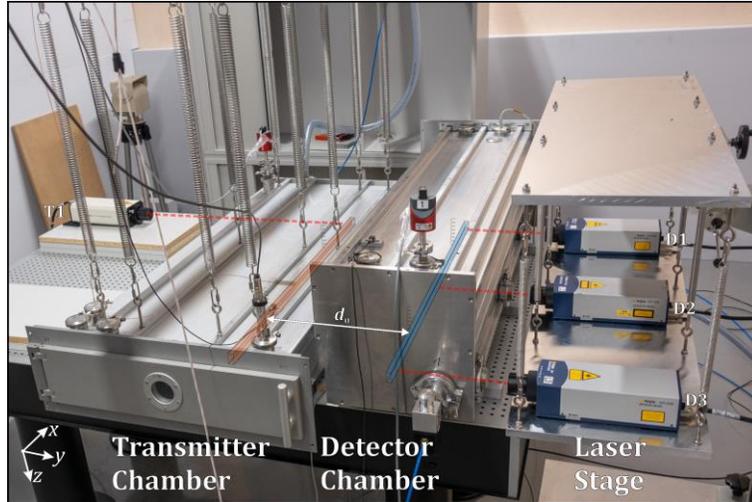

**Figure 2: Illustration of the measurement setup.** The transmitter beam (red drawing), hanging on springs attached at the nodal points of the first bending mode, is located inside the transmitter chamber. The chamber itself is likewise hanging from springs attached to a carrier bar movable in *y*-direction. The vibration amplitude of the transmitter is measured using a single point laser interferometer (T1). The detector beam (blue) is similarly mounted, whereas the detector chamber is placed on an anti-vibration table, isolating the beam from the environment and minimizing transmission of non-gravitational forces. The detector movement is measured using three laser vibrometers (D1-D3), positioned on a separate stage, likewise isolated via springs. By moving the transmitter chamber, the distance $d_0$ between the beams can be varied. The whole setup is located in an underground laboratory, providing excellent temperature stability and minimal seismic noise. The accelerometers measuring the chamber movement are not visible in the picture.

To maximize the amplitude of the detector beam, one shall strive for a large transmitter mass, low frequencies, small distance and a high Q of the detector oscillator, cf. Eq. (2). Therefore, we selected a detector beam made of titanium of 1 m length and a rectangular cross section of 17 mm x 8.5 mm. All experimental details are given in the supplementary material. The beam is hanging on two strings that are attached at the nodal points of the first bending mode to minimize both additional damping and the transmission of forces from the supporting structure. An illustration of the setup is shown in Fig. 2, containing a picture of the setup and illustrations of both the transmitter and detector beam.

The detector motion is measured using three laser interferometers (D1-D3 in Fig. 2) placed on a separate, spring-suspended platform. The laser beams point horizontally onto the beam surface where they are reflected by small patches of retroreflective tape. The use of three measurement positions allows to extract the bending motion from signals that may also contain the rigid body motion of both the detector beam and the laser stage (cf. Methods).

The transmitter beam is made of tungsten with dimensions 1 m x 20 mm x 10 mm and a total mass of 3875.6(1) g. A piezoelectric transducer with a counter-mass at the opposite end is mounted at the centre of the beam to excite the bending motion with minimal reaction forces acting on the supports. The system is designed such that the first bending resonance frequency is as close as possible to the bending resonance of the detector beam (ca. 42.65 Hz). The transmitter beam is hanging on two springs that are attached at the nodal points of the first bending mode. The beams displacement is measured by another laser interferometer (T1 in Fig. 2).

Besides a maximization of the gravitational forces and the resulting detector motion, the prevention of any excitation or signal coupling other than gravity is of utmost importance. Great care is taken to avoid any system resonances near the detector resonance. Also, both beams have been placed in separate aluminium vacuum chambers to avoid acoustic coupling effects. Further, the chamber containing the detector was placed on a vibration isolation table, while the transmitter chamber itself is hanging on springs from a movable bar attached to a solid frame with high damping. Accelerometers mounted on both chambers give information about unwanted movement of the chambers.

Due to the high Q-factor of the detector beam of $Q_d \approx 35000$, the time until a steady state vibration is reached is quite high (time constant $\tau \approx 260$ s). Further, long-time averaging will be applied during evaluation to increase the measurement resolution. Hence it must be ensured that the vibration conditions, especially the resonance frequency of the detector beam, remain stable during the measurement period. Therefore, the whole setup has been placed in an underground laboratory in the Swiss Alps where a very stable temperature can be guaranteed, resulting in an average temperature span of ca. 8 $10^{-4}$ °C per measurement point (ca. 50 min). Additionally, the location shows minimal external disturbances comparable to a VC-G facility[27].

Further details on the experimental setup and the isolation measures can be found in the supplementary material.

During an experiment run, the piezoelectric transducer is sequentially excited with a 14 $V_p$ sinusoidal signal at frequencies around the detector bending resonance. Each frequency is held for 50 min and the detector and transmitter velocities are measured by the laser vibrometers. Both transmitter and detector chamber are evacuated to a pressure of 1 mbar before every new frequency, yielding a constant pressure within 1% during 50 min due to minor leakage. Therefore, it is guaranteed that the dynamic properties of the detector remain unchanged throughout the whole measurement.

To extract amplitude and phase at the frequency of excitation only, the outputs of the laser interferometers are fed into individual lock-in amplifiers, thus achieving a velocity amplitude resolution of 0.16 nm/s. The data from the last 25 min of every frequency step is averaged and used for the data evaluation, further improving the resolution.

Figure 3 shows the result of an exemplary measurement run (30 frequencies, frequency resolution Δf = 0.126 mHz, total duration ca. 31 h). Due to the relatively high damping of the transmitter beam ($Q_t \approx 600$), the frequency spectrum of the transmitter beam reveals a nearly constant bending vibration of ≈ 94 mm/s amplitude as shown in column (a). Column (b) shows the frequency spectrum of the detector's bending amplitude around the bending resonance, derived from the velocities at three measurement positions along the detector beam, thereby eliminating pendulum motions (cf. Methods).

The spectrum shows a clear resonance at 42.651 Hz corresponding to the considered bending mode. By taking the ratio between excitation and response velocity, the transfer function $v_d/v_t$ can be derived, as shown in column (c). According to Eq. (2), the amplitude ratio at resonance $v_{d,0}/v_{t,0}$ gives a measure of the gravitationally induced excitation. Note that the phase at resonance is close to -90° and not +90°, as Eq. (2) would indicate, since transmitter and detector laser measure the velocity in opposite directions.

Since the bending resonance can be approximated as a single-degree-of freedom (SDOF) oscillator (provided that no other resonances are close-by), the well-known equation of such systems can be fitted to the measured transfer function. As a result, an estimate for the detector's resonance frequency, amplitude ratio at resonance, Q-factor and resonance phase shift is obtained (cf. Methods). $Q_d$ was also confirmed alternatively by a piezoelectric excitation of the system.

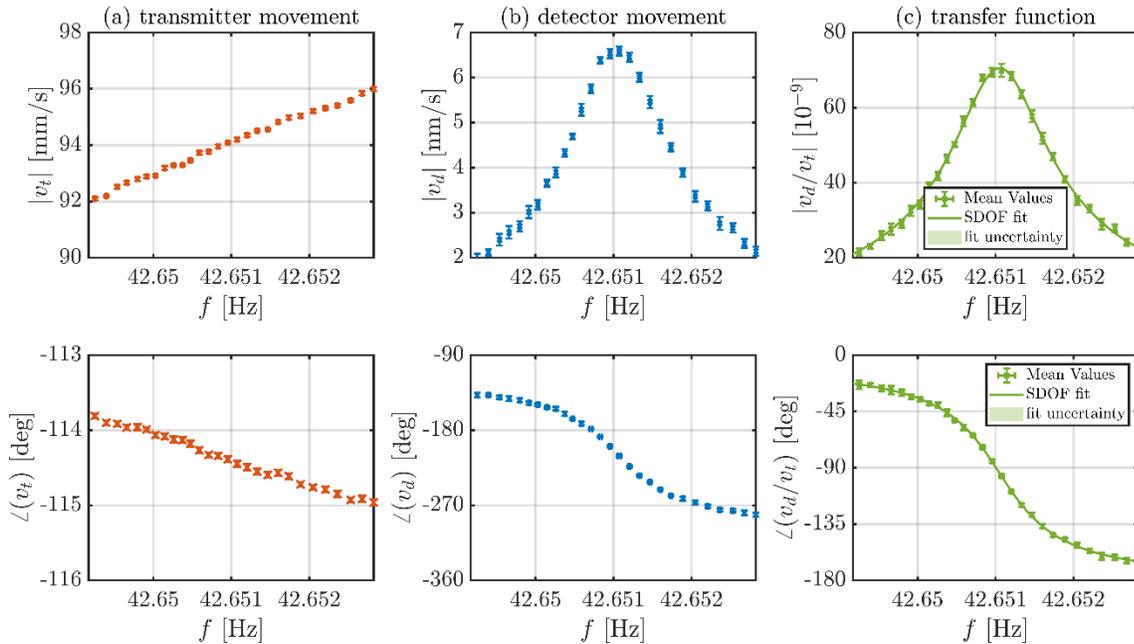

**Figure 3: Result of a measurement run.** Transmitter and detector frequency spectrum (Top row: amplitude; Bottom row: phase) evaluated from subsequent excitation at 30 discrete frequencies around the first bending resonance of the detector beam at a beam distance of $d_0$ = 59 mm. (cf. run 18-Mar-21-1 in Fig. 5). Each frequency measurement took 50 min, of which the data points of the last 25 min are averaged. Mean values and errorbars are depicted in the plot. **a**, Transmitter velocity spectrum: quasi-constant excitation with ca. 94 mm/s amplitude. **b**, Detector bending velocity spectrum extracted from three measurement positions. **c**, Transfer function spectrum calculated from (a) and (b). The fitting of a single-degree-of-freedom (SDOF) transfer function allows to extract the parameters for this measurement: resonance frequency $f_0 = \omega_0/2\pi$ = 42.65105015(77) Hz, Q-factor $Q_d$ = 3.5827(46) $10^4$, amplitude ratio at resonance $v_{d,0}/v_{t,0}$ = 70.14(13) $10^{-9}$ and resonance phase shift $\varphi_0$ = -93.965(73) deg. The coefficient of determination of the fit is $R^2$ = 99.64 %.

# Measurements and results

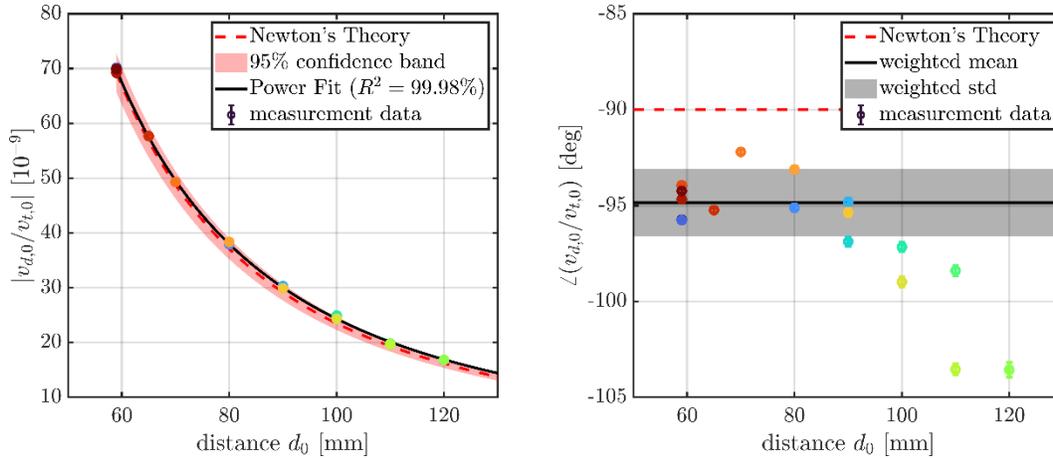

**Figure 4: Gravitationally induced bending motion.** Amplitude ratio between detector and transmitter beam (left) and corresponding phase difference (right) at resonance as a function of the beam distance $d_0$. Each data point shows the SDOF fit result and according confidence interval obtained from a frequency spectrum measurement. The amplitude curve shows a power-law behaviour with exponent -1.99(1) and $R^2$ = 99.98% (black line in left plot). The values lie well within the confidence band (red area) of the theoretical prediction (red dashed line) that comprises statistical and systematic uncertainties, as summarized in Extended Data Table 1. The inverse-variance weighted mean phase shift of (-94.8 ± 1.7) deg (black line in right plot) is a bit lower than the theoretical value of -90 deg, suggesting minor error contributions. The colour of the data points is the same as in Fig. 5.

Over a period of four weeks, 18 measurement runs like the measurement presented in the previous chapter (cf. Fig. 3) were conducted at different beam distances.
The results of the amplitude ratio at resonance $v_{d,0}/v_{t,0}$ are depicted in Fig. 4. Each data point corresponds to a 22-point frequency spectrum, where resonance amplitude and phase have been extracted by fitting the transfer function of a SDOF oscillator. The amplitude curve shows a clear power-law behaviour with exponent -1.99(1) and $R^2$ = 99.98%. The excellent agreement with Newton's theory implicates that the detector motion is indeed caused by gravitational coupling and not due to other effects such as mechanical or acoustical coupling.
The values lie well within the tolerance band of the theoretical prediction that comprises statistical and systematic uncertainties, as summarized in Extended Data Table 1. The inverse-variance weighted mean phase shift of (-94.8 ± 1.7) deg is a bit lower than the theoretical value of -90 deg, suggesting a minor error contribution. With distance the deviation of the resonance phase from the theory increases, while the standard error remains rather unchanged. This may indicate a systematic error due to remaining crosstalk, which becomes more prominent with decreasing gravitational signal.
Since temperature and pressure barely change, the Q-factor of the detector beam was estimated by an inverse-variance weighted mean of all measurement results, resulting in $Q_d$ = 3.595(85) $10^4$. Further, the resonance frequency of the detector was determined to be $f_0 = \omega_0/2/\pi$ = 42.648925(55) Hz at 11.5°C, being linearly dependent on the temperature with a coefficient of $\alpha$ = -0.01834(56) Hz/°C.
Finally, *G* was estimated from each measurement result using Eq. (2) and the theoretical model. In Fig. 5 the single results are depicted as mean value and standard deviation (coloured patches). As an overall result, the inverse-variance weighted mean of the individual measurements of *G* (black line) yields an estimate of $G^*$ = 6.821(71) $10^{-11}$ m$^3$kg$^{-1}$s$^{-2}$, which is only about 2.3% higher than the CODATA 2018 value[11]. The overall confidence band (black dashed line in Fig. 5) represents the extended measurement uncertainty (k = 1.96) based on the statistical and systematic uncertainties summarized in Extended Data Table 1.
Due to the observed deviation from theory, a slight systematic error might be present, which is suspected to be a non-gravitational, mechanical coupling. During these experiments we measured a small movement of the detector chamber, typically about 0.25 nm/s at the frequency of excitation, which might be responsible.

Based on a power balance, we can also compute the near field gravitational energy flow between transmitter and detector. At steady state incoming energy at the detector is dissipated according to the Q-factor of the detector. If all this energy is attributed to gravitation, this yields a gravitational power of 2.64 $10^{-20}$ W for the measurement run depicted in Fig. 3.

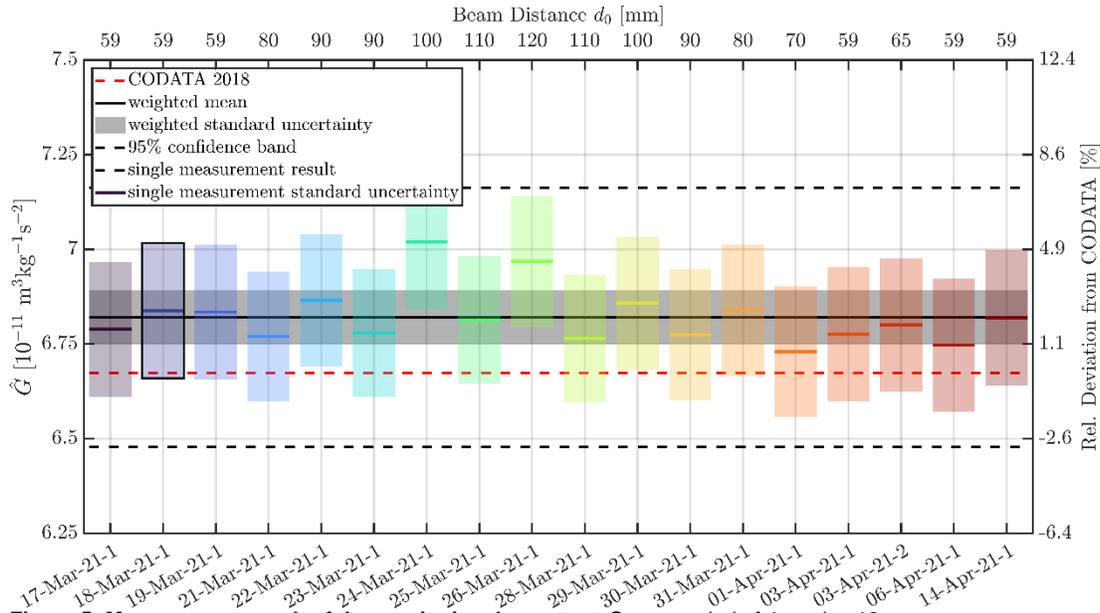

**Figure 5: Measurement result of the gravitational constant.** Over a period of 4 weeks, 18 measurement runs were conducted at different beam distances $d_0$ (upper x-axis). From the fitted resonance amplitude, estimates for the gravitational constant were derived, using a fixed detector beam Q-factor of 3.587(88) $10^4$, obtained as weighted mean value from all individual measurement results. Using the theoretical values for the corresponding distance, each measurement yields an estimate for *G* (coloured lines), together with a standard uncertainty (coloured boxes). The black framed box corresponds to the measurement shown in Fig. 3. An inverse-variance weighted mean (black line and black shaded area) yields an overall estimate of *G*\* = 6.821(71) $10^{-11}$ m$^3$kg$^{-1}$s$^{-2}$, which is about 2.3% higher than the CODATA 2018 value (red dashed line). The plotted 95% confidence band (black dashed line) includes systematic and statistical contributions as summarized in Extended Data Table 1. The colour of the data points is the same as in Fig. 4.

# Discussion and outlook

We have demonstrated the dynamical gravitational coupling between two beams vibrating in their first bending resonance around 42 Hz both experimentally and theoretically. To our knowledge, this is the first dynamical gravity experiment in this frequency range where a full quantitative comparison between measurement and theory was successfully done. Experimental investigations show excellent first results where we obtained a gravitationally induced detector velocity amplitude of about 6.6 nm/s. Due to thorough acoustical, mechanical, and electrical isolation, a temperature-stable environment and lock-in measurement technique we were able to measure said motion with a signal-to-noise ratio up to 100 at a measurement time of 50 min per data point.

By varying the beam distance, we observed an inverse square law of the detector amplitude. Together with the excellent agreement with the theoretical model, this implicates the gravitational character of the coupling forces.

The comparison with the theoretically expected results demonstrated not only the feasibility of the methods but also promising accuracy. We were able to estimate the gravitational constant to $G = 6.821(71)\ 10^{-11}$ $m^3 kg^{-1} s^{-2}$ which is about only 2.3% higher than the CODATA 2018 value[11].

We demonstrated numerous advantages of the method, such as fast measurement time, easy variation of the distance, high output amplitudes and the decoupling from static gravitational fields. Thanks to the dynamic nature of the interaction, it was even possible to determine the energy transmitted from transmitter beam to the detector beam, revealing a $d_0^{-4}$ behaviour of the energy flow.

However, many things can be improved. This includes measures that increase the detector amplitude, such as the use of a detector material with higher Q, such as quartz glass, or higher excitation amplitudes. Since the measured $G$ is a bit higher than the CODATA value, an additional systematic error seems to be present, which is suspected to be a slight mechanical coupling. Elimination of said coupling, e.g. by an active vibration isolation system might increase the accuracy of the results considerably.

The results still have an uncertainty that is primarily affected by the relatively large uncertainty of the distance measurement and the detector beam quality factor $Q_d$. Improvements of the setup will also decrease the uncertainty of these parameters.

The combination with well-established resonance control can improve measurement speed and accuracy.

Motivated by the work of Sinsky/Weber[20], future work can also contain a non-resonant transmitter such as one or several rotating bars, which enables to use higher harmonics of the dynamic gravitational forces.

We think that this work shows the high potential of fully characterized dynamical gravitational experiments. The approach can be transferred to other continuum vibrational systems in micro- and macroscale thus opening a completely new field of gravitational interaction experiments, leading to new insights in this field.

Besides a more accurate determination of $G$, our next goals include topics such as the investigation of the frequency dependency of $G$, the inverse square law or gravitational shielding at high frequencies.

# Methods

## Theory

### Point mass model

The equation of motion of the detector mass as shown in Fig. 1a is given by

$$m_d \ddot{u}_d + \delta \dot{u}_d + c u_d = -F_G = -G \cdot \frac{m_d m_t}{d^2} \tag{3}$$

where $\delta$ is the damping constant, $c$ the spring constant, $u_d$ the detector mass displacement with respect to its reference position, $d$ the distance of the masses $m$ and $F_G$ the gravity force according to Newton's law of gravitation[1,2]. Indices $_t$ and $_d$ indicate the element being a part of the transmitter or detector, respectively. A superposed dot denotes the time derivative. We assume low damping $\delta \ll 2\sqrt{(m_d c)}$. The transmitter mass is oscillating with amplitude $u_t'$ and angular excitation frequency $\omega$, hence the distance between the masses changes periodically with $d = d_0 - u_t' e^{i\omega t}$. Assuming $\varepsilon = u_t'/d_0 \ll 1$ allows to approximate the force with a Taylor expansion around $\varepsilon = 0$ as

$$F_G \approx \frac{G m_d m_t}{d_0^2} + \frac{2 G m_d m_t}{d_0^3} u_t' e^{i\omega t} + O(\varepsilon^2) \tag{4}$$

representing a static component, a dynamic part at frequency $\omega$ and negligible terms of higher order in $\varepsilon$. Instead of using the Taylor approximation, the full nonlinearity can be considered by computing the Fourier series coefficient $c_1$ at the frequency $\omega$, i.e.

$$F_{G,\omega} = c_1 \cdot e^{i\omega t} \; ; \; c_1 = \frac{2\pi}{\omega} \int_0^{\frac{2\pi}{\omega}} F_G(t) \cdot e^{-i\omega t} dt \tag{5}$$

Introducing the angular resonance frequency $\omega_0 = (c/m_d)^{1/2}$ and quality factor $Q_d \approx (c m_d)^{1/2} \delta^{-1}$, the detector mass displacement is given by the well-known solution of Eq. (3)[3]

$$u_d = -\frac{F_G}{m_d} \cdot \frac{1}{\omega_0^2 - \omega^2 + \frac{i\omega\omega_0}{Q_d}}. \tag{6}$$

When exciting at resonance ($\omega = \omega_0$), the maximum displacement amplitude of the detector is obtained, given by Eq. (1). Because $Q_d$ is of the order of $10^4$ and higher, the static component and higher order terms can be neglected in the vicinity of $\omega_0$.

### Beam model

Assuming continuous bodies as shown in Fig. 1b, the gravitational interaction can be formulated in terms of continuum mechanics, which yields the gravitational force in y-direction between two infinitesimal mass elements $dm$ of detector and transmitter beam given by

$$F_{Gy}^{(dm_t - dm_d)} = -G \frac{\mathbf{r} \cdot \mathbf{e}_y}{\|\mathbf{r}\|^3} dm_t dm_d \; ; \; \mathbf{r} = \mathbf{r}_{dm_d} - \mathbf{r}_{dm_t} \tag{7}$$

with the position vectors $\mathbf{r}_{dm}$ of the mass elements $dm$ and the unit vector $\mathbf{e}_y$ in y-direction. The position vectors include amplitude ($u_t'$ and $u_d'$, respectively), frequency $\omega$ and shape of the beam's centre line displacements that can be derived from the well-known Euler-Bernoulli beam theory using free-free boundary conditions[4]. The use of this linear theory for the first bending mode is justified because the beams are slender (thickness/wavelength $h/\lambda \approx 0.01$, width/wavelength $w/\lambda \approx 0.01$) and the displacement amplitudes $u$ are small ($u \ll h, w$). Note that the mode shapes of detector and transmitter slightly differ due to the additional mass of the piezo-transducer and counter-mass mounted on the transmitter beam. Further, the $x$-displacements (cf. Fig. 1b) are neglected as they are of the order of $(h/\lambda) \cdot du/dx_t$.

Consequently, the gravitational force distributed along the central axis of the detector beam can be derived by integrating over the transmitter beam volume $V_t$ and detector beam cross-sectional area $A_d$ which yields the force per unit length in $y$-direction

$$F_{Gy}(x_d, t) = -G\rho_d\rho_t \int_{(V_t)} \int_{(A_d)} \frac{\mathbf{r} \cdot \mathbf{e}_y}{\|\mathbf{r}\|^3} dV_t dA_d \approx -G\rho_d\rho_t \left( f_{Gy,0}(x_d) + f_{Gy,1}(x_d) e^{i\omega t} \right) \tag{8}$$

as a function of the detector coordinate $x_d$, where $\rho$ denotes the material density, assuming a homogeneous mass distribution in each beam. Similar to the point mass model, a Taylor series approximation around $\varepsilon = u_t'/d_0 = 0$ can be applied, yielding a static part $f_{Gy,0}$, a dynamic part $f_{Gy,1} e^{i\omega t}$ containing the frequency of excitation and higher order terms $O(\varepsilon^2)$. Alternatively, a Fourier series can be used if the nonlinearity of the gravitational force must be taken into account (cf. Eq. (5)). For the small vibration amplitudes used in the experiment, however, the two approaches yield very similar results with a difference of less than 0.5%.

For $F_{Gy}$ we neglect the influence of the very small, gravitationally induced detector motion as it has a negligible influence ($u_d' < 10^{-8} \, u_t'$).

The solution of the detector motion can be found using the eigenfunction expansion method[4], where a particular solution of the form

$$u_d(x_d, t) = u_T(t) \cdot U_T(x_d) + u_R(t) \cdot U_R(x_d) + \sum_{n=1}^{\infty} u_{b,n}(t) \cdot U_{b,n}(x_d) \tag{9}$$

is considered. $U_T$ and $U_R$ denote translational and rotational rigid body movements with corresponding contribution factors $u_T$ and $u_R$, respectively, while $U_{b,n}$ are the normalized eigenfunctions ($U_{b,n}(0) = 1$) and $u_{b,n}$ the corresponding modal coordinates[4]. When the excitation frequency is close to a particular resonance frequency (in our case the first bending mode resonance frequency $\omega_0 = \omega_{b,1}$) it can be shown, based on the orthonormality relation of the eigenfunctions and the symmetry properties of the force field, that the contributions of all other modes are negligible. This applies as well for the rigid body motions. The amplitude of the corresponding modal coordinate is thus found by the solution of the first order ODE

$$\ddot{u}_{b,1} + \frac{\omega_0}{Q_d} \dot{u}_{b,1} + \omega_0^2 u_{b,1} = -\frac{4G\rho_d\rho_t}{m_d} \int_0^{l_d} U_{b,1}(x_d) f_{Gy,1}(x_d) e^{i\omega t} dx_d \tag{10}$$

where the modal damping is specified with the quality factor $Q_d$ of the detector beam. When steady state is reached, the detector motion is obtained from the particular solution of Eq. (10), i.e.

$$u_d(x_d, t) = u_d' \cdot U_{b,1}(x_d) \cdot e^{i\omega t} \tag{11}$$

where $u_d'$ is the gravitationally induced vibration amplitude that will be measured by the detector lasers.

The integral in Eq. (10) can be simplified by a transformation to dimensionless variables (denoted with bar accent in Eq. (12)). Then the detector displacement amplitude is obtained as given in Eq. (2), which contains the frequency dependence as given in Eq. (6), the easy to measure properties of the experiment and a 6$^{th}$ order integral over all spatial dimensions. The integral can be computed with arbitrary precision using a numeric computing environment (we used Maplesoft Maple 2020).

$$u_d' = u_t' \cdot \frac{-Gm_t}{\omega_0^2 - \omega^2 + \frac{i\omega\omega_0}{Q_d}} \cdot \underbrace{4 \int\limits_{(\bar{V}_t)} \int\limits_{(\bar{V}_d)} \bar{U}_{b,1} \bar{f}_{Gy,1}(\bar{x}_d) d\bar{V}_d d\bar{V}_t}_{\Gamma(d_0, \text{beam coordinates}, \text{beam dimensions})} \quad (12)$$

The gravitational influence of other moving masses such as the transducer and the counter mass are calculated likewise, resulting in an additional excitation force. For the setup and parameters presented here, the integral term including all exciting masses can be well approximated by $\Gamma(d_0) \sim d_0^{-2}$.

## Limitations and prospects of the model

In this chapter some limitations of the theoretical model shall be discussed briefly.

Regarding the approximation methods, the Taylor approximation is favourable, since only a 6$^{th}$ order integral must be solved. However, if $u_t'/d_0 \ll 0$ does not hold any more, a Fourier approximation will be the better choice. Generally, the Fourier approximation is to be preferred if computational power is not limited, since it considers the full non-linearity of the gravitational force field.

For small distances the excitation force might contain significant amplitudes at higher harmonics of the excitation frequency that might excite other vibration modes of the detector beam. Although the lock-in measurement technique enables to extract the motion at one specific mode/frequency, said harmonics might add a certain error to the results. A multi-modal analysis, i.e. additional measurements at higher harmonics could be used to even better quantify the gravitational coupling[5].

## Energy considerations

When steady state is reached, the energy stored in the vibration of the detector beam can be calculated from the kinetic energy at maximum beam velocity, i.e.

$$E_{\text{stored}} = \frac{1}{2} \rho_d A_d \omega^2 u_d'^2 \int\limits_0^{l_d} U_{b,1}(x_d)^2 dx_d \quad (13)$$

For a bending amplitude of $u_d' = 24.6$ pm (resonance amplitude of run 18-Mar-21-1, cf. Fig. 3b) this yields a stored energy of 3.54 10$^{-18}$J.

According to the definition of the Q-factor as $2\pi$ times the ratio between the stored energy and the energy dissipated per cycle[6], the dissipated energy can be calculated using the measured Q-factor of the detector beam ($Q_d = 3.595 \; 10^4$).

If we assume that the only source that introduces energy into the detector beam is the gravitational force, the time averaged energy flow from transmitter to detector beam must be equal to the energy dissipated in the detector beam in order to maintain the vibration. For our experiments, we obtain a gravitational energy flow of 2.64 10$^{-20}$ J/s during resonance excitation in

run 18-Mar-21-1.

Consequently, the gravitational energy flow decreases according to $d_0^{-4}$ since the detector bending amplitude is depending on approximately $d_0^{-2}$.

## Experimental setup

Because of the weak nature of gravitational interaction great care must be taken to minimize undesired effects. In particular, one has to make sure that no other vibrational modes exist in the vicinity of the excitation frequency. Because of the high Q factor of the used detector mode of > $10^4$, the desired motion is then strongly amplified in comparison to all other contributions. To maximize the amplitudes, a low frequency, small distance $d_0$ and large transmitter mass $m_t$ are desired. At the same time dimensions compatible with an easily accessible laboratory set up and vacuum chambers must be used, which make a thorough and contactless investigation of non-gravitational crosstalk effects possible. The setup is fully remote controlled and automated using LabVIEW$^{TM}$ in order to avoid disturbances by the operators.

### Transmitter beam

The transmitter beam has a length of 1.0004(1) m with a rectangular cross-section of 20.70(1) mm x 10.05(1) mm. To have different resonance frequencies for the bending modes in *z* and *y* direction, respectively, the area moment of inertia must be different for the two directions. For precise machinability a rectangular cross-section was chosen. The mass of the transmitter is 3875.6(1) g, the Q-factor of the first bending mode (internal + external damping) is ca. 680 at low amplitudes. The transmitter resonance was measured to be 42.58 Hz. Due to the relatively small Q-factor and a small transmitter nonlinearity, sufficient vibration amplitude is achieved in the vicinity of the detector's resonance frequency of 42.651 Hz, making transfer function measurements possible. The beam has been cut from 99.95 % pure tungsten sheet material (ASTM B760). It is hanging on taut springs of 44 mm length (spring constant ca. 0.94 N/mm) attached to the nodal points of the first bending mode to minimize transmission of forces to the transmitter chamber. Due to the inertial excitation by piezoelectric elements, no external forces are applied. The nodal points are calculated using the Euler-Bernoulli beam theory, including an attached mass at the centre of the beam of 690 g. Small nonlinearities due to relatively high amplitude of the transmitter can be seen in the transmitter's transfer function. However, the displacement function is assumed to be negligibly influenced as the ratio $u_t'/l_t < 10^{-3}$.

### Detector beam

The detector beam is made from grade 2 titanium (3.7035) and has a length of 1.0000(1) m with a rectangular cross-section of 16.97(1) mm x 8.49(1) mm. The mass of the detector beam is 647.72(2) g. Titanium has been selected because of its high Q factor amongst metals[7]. The beam is isolated from the environment by hanging on ethylene propylene diene monomer (EPDM) rubber strings of 3 mm diameter and 268 mm length, glued to the nodal points of the first bending mode. For a better decoupling, a mass of 7.1 g has been placed in the centre of each rubber wire. Since the EPDM strings showed a certain creep behaviour, they have been loaded for several weeks with a mass similar to the detector beam until the length was stable. During the experiments, an elongation of the wires of less than 0.5 mm per week was monitored, resulting in an increase of the beam distance $d_0$ of < 0.0035% at $d_0$ = 59 mm.

### Excitation

For the excitation of the transmitter beam a preloaded piezoelectric transducer (PI P-843.60) is mounted at the centre of the beam. To increase the excitation force acting on the beam, a counter-mass of 592 g is mounted at the opposite end of the actuator, yielding a dynamic force of 0.566(3) $N_p$ at an excitation amplitude of 14 $V_p$ at 42.651 Hz (PI E-505 amplifier).

The excitation signal is provided by a SRS FS740 time and frequency system, which acts as a time base for the whole experiment. The device has a phase noise < -130 dBc/Hz and frequency stability < $10^{-12}$ (Allan deviation[8]),

### Distance

The distance between the beams was adjusted by moving the transmitter chamber that is hanging on 12 springs from two stiff bars mounted on two parallel, synchronized linear motors (Bosch Rexroth CKK-200-NN-1 with IndraDrive HCS01 control) which can move with µm precision. However, the initial distance $d_0$ could not be maintained precisely yet, resulting in a systematic distance uncertainty of 0.5 mm.

### Laboratory location/seismic noise

In order to obtain highest temperature stability together with minimal seismic noise, the measurement equipment was installed in a former underground military facility in the Swiss alps. Prior to the installation, the room was characterized in terms of its vibration characteristics by an external company, using a Syscom MR3000TR/MS2003+ triaxial geophone. The measurement revealed excellent conditions with floor vibrations that undercut the VC-G criterion[9] (< 0.8 µm/s) by about a factor of 20. The passive temperature stability of the room itself is excellent with temperature drift of < 0.04 °C/h, two days after persons have left the rooms.
The experiment is distributed across three rooms: the measurement room containing the vacuum chambers, beams, sensors and a minimum of other equipment, the control room where all computers and controls, as well as remote connectivity are located and the pumping room containing the vacuum pumps and a dehumidifier.

### Velocity measurement

The movement of the detector beam was measured with three separate laser interferometers (Polytec OFV-505 laser head + Polytec OFV-5000 controller with VD-06 decoder) that are placed on a separate mount (cf. Fig. 2), connected by weak springs to a solid support. The laser beams point horizontally and orthogonally (deviation ± 1°) onto the beam surface where they are reflected by 10 mm x 10 mm patches of retroreflective tape (3M Scotchlite 7610) sticking on the beam's surface. The mass of the tape is considered negligible (0.02 g per patch). The laser vibrometers provide a velocity resolution of ca. 10 nm $s^{-1}$ $Hz^{-1/2}$, which yields a measurement resolution of ca. 0.37 nm/s using an $8^{th}$ order, 1.37 mHz lowpass lock-in amplifier with a time constant of 31.32 s (Zurich Instruments MFLI). The inputs of the lock-in amplifiers are attenuated by a factor of 4 in order to prevent overload damage of the 3 V input stages of the MFLI. This could happen if the lasers lost focus, which results in voltages of 10 V (an extremely rare event).
To quantify the resolution and quality of the velocity measurement, calibration measurements have been conducted prior to the experiments. For this purpose, one of the laser heads was adjusted to point onto a small piezoelectric transducer (10 mm x 10 mm x 1 mm) glued onto a solid steel block. Apart from that, the measurement setup remained unchanged. The transducer was driven with different excitation amplitudes at 42.6 Hz, resulting in velocities up to 10 nm/s. Measurements with zero amplitude revealed a noise level in the range of 0.16 nm/s (50 points of averaging), while standard deviations of signals in the range of the highest amplitudes measured in this paper (ca. 7 nm/s) showed 0.1 nm/s. These numbers are even better than the aforementioned expected resolution. Consequently, signals > 1.6 nm/s can be measured with a SNR > 10. This can be further improved by longer averaging and/or larger lock-in time constants. Considering a fixed frequency of 42.6 Hz, 1.6 nm/s corresponds to a displacement resolution of 0.6 pm since $v = \omega\, u$.
Since the amplitudes are small in these calibration measurements, the behaviour of the piezo is linear and an amplitude sweep was conducted to reveal nonlinear effects of the measurement chain. The amplitude range was selected to be similar to the measurement range in the experiments. A linear fit of the excitation amplitude vs. velocity data points revealed a non-linearity < 2 %$_{FSO}$, where full scale output (FSO) is FSO = 8 nm/s. Note that this measurement also contains nonlinearities in the excitation chain (i.e. the driving of the piezo), which is different from the excitation in the experiments.

## Vibration isolation and measurement

*Acoustical isolation*

To avoid the transmission of forces due to sound waves, each chamber was evacuated with a vacuum pump (Edwards nXDS 10i vacuum scroll pump) via remote controlled valves. Pressure evacuation was done before every frequency step, whereas no data was acquired during evacuation. The pressure increase of the chambers for pressures > 0.5 mbar was measured to be 22 µbar/h for the detector chamber and 155 µbar/h for the transmitter chamber.

*Mechanical isolation*

Besides the spring suspension of both beams and the transmitter chamber, the detector chamber was placed on an aluminium vibration isolation table (Opta HDT 200 anti-magnetic table with DMT 1400 base).

*Electrical isolation*

To minimize electrical coupling of the excitation signal into the measurement chain, e.g. via ground loops containing the excitation frequency, all relevant devices have been connected in a star ground layout to the main power source. Any high-power sources were disconnected during the measurement. Further the detector beam as well as the vacuum chambers were electrically grounded to avoid patch charges and creating Faraday cages.

*Mechanical crosstalk measurement*

Despite the thorough vibration isolation, minor transmission of forces and signals were detected at the measurement frequency. We assume these effects to be responsible for the remaining uncertainties and deviations in the measurement results.
During the measurements presented in this article, we measured a movement of the transmitter chamber of 0.82(2) µm/s in y-direction, and 0.15(2) µm/s in z-direction using a Bruel&Kjær 4535-B-001 triaxial accelerometer. A certain movement of the transmitter chamber, however, is not critical if it does not excite the detector beam in its bending mode (cf. Theory supplement).
The movement of the detector chamber was also measured using two Kinemetrics EPI ES-T FBA triaxial accelerometers placed on top of the chamber. These accelerometers provide a resolution of ca. 1.4 pm/s when combined with an 8$^{th}$ order, 1.37 mHz lowpass lock-in amplifier. Typically, we measured detector chamber velocities of ≈ 0.25 nm/s in y and ≈ 0.4 nm/s in z-direction at the frequency of excitation.

## Post processing/data evaluation
### Offset correction

To eliminate possible errors due to a frequency independent, time constant offset value of the measurement chain, an additional measurement at a frequency far away from the detector's resonance has been made prior to each measurement run (here we used 40.44 Hz). Thus, the signal level with active piezo excitation can be determined and subtracted from the measurement output, eliminating also electrically coupled signals produced by the amplifier of the piezo excitation. The typical offset value measured by the lock-in amplifiers was in the range of 0.3 nm/s.

**Bending amplitude extraction**

Due to the detector beam spring suspension, rigid body movements occur as pendulum movements with resonance frequencies around 1 Hz, velocity amplitudes up to 6 µm/s and very high time constants. Once excited by additional disturbances they might remain for a long time and are therefore still observed in the measurements. Despite their low resonance, they may have a contribution to the velocities measured at the frequency of excitation as well.

Therefore, the measurement setup uses three lasers to extract the amplitude of the first bending mode from the combined motion

$$u_d(x_d) = \left( u'_T + u'_R \left( \frac{2}{l_d} x_d - 1 \right) + u'_{b,1} U_{b,1}(x_d) \right) e^{i\omega t} \qquad (14)$$

measured by the lasers. The first bending mode shape $U_{b,1}$ is described by the well-known normalized solution of an Euler-Bernoulli bending beam with free-free boundary conditions[4]. Using the measured displacement/velocity at the three measurement positions $x_d$ = [0.005 m, $l_d$/2, $l_d$ - 0.005 m] of the detector beam, the amplitudes $u_T'$, $u_R'$ and $u_{b,1}'$ can be calculated by solving the system of equations that results from Eq. (14).

Typical rotational and translational movement at the measurement frequency were in the range of < 1E-10 m/s.

**Temperature correction**

From preliminary experiments a linear temperature dependency of the detector's resonance frequency was found (temperature coefficient of $\alpha$ = -0.01834(56) Hz/°C).

Hence, since the temperature might slightly change during a measurement, the resonance frequency changes as well, thus producing a deviation of the frequency response from the expected SDOF behaviour (see Section "SDOF Fit").

Assuming that the temperature behaviour is equal for all frequencies around the resonance, one can compensate for this effect by subsequently shifting the excitation frequencies with respect to the mean temperature of the measurement.

However, due to the very stable temperature inside the detector chamber ($\Delta T$ < 0.005 °C per hour), this effect had a negligible influence on the overall results.

**SDOF fit**

The quantification of the gravitational coupling between the two beams requires the knowledge of the response amplitude of the detector and its vibrational properties. It has been shown that the frequency response of the detector beam around the resonance can be modelled by the response of a single-degree-of-freedom (SDOF) oscillator (cf. Section Theory). Thus, the theoretical equation of the transfer function *TF* of a SDOF oscillator

$$TF(\omega) = \frac{A_0 \cdot \frac{\omega_0^2}{Q_d}}{\omega_0^2 - \omega^2 + \frac{i\omega\omega_0}{Q_d}} \cdot e^{i\phi_0} + XT \qquad (15)$$

is fitted to the measured transfer function that is obtained from the detector bending velocity and the velocity of the transmitter beam (cf. Fig. 3). $A_0$ denotes the amplitude ratio at the resonance frequency $\omega_0$, corresponding to $|u_{d,0}/u_{t,0}|$. $Q_d$ is the Q-factor of the detector's first bending mode. As additional parameters the phase at resonance $\phi_0$ is introduced to account for eventual deviations from the theoretical prediction ($\phi_0$ = -π/2). Further, a (complex) offset *XT* is introduced to include frequency independent crosstalk. The fitting is carried out with MATLAB using a Levenberg-Marquardt non-linear least squares solver.
In order to assess the uncertainties of the individual parameters, a bootstrapping method has been applied that uses 100 individual fits based on randomly sampled data (with replacement)[10].

## Uncertainty estimation

Although a proof of principle state, an estimation of the measurement uncertainties of our experimental results is presented in this section. It shall be noted, however, that this article presents a novel and fully characterized experiment in uncharted territory with the benefits of a dynamical measurement, hence the assessment of the uncertainty should be regarded both as preliminary and an attempt to stimulate further ideas around this interesting aspect.
In estimating the uncertainty associated with measuring $G$, Eq. (2) is used to assess the influence of the different variables. Therefore, one needs to evaluate the contribution of the function Γ to the overall error as well. This is a rather challenging task since the function depends on several parameters in a highly non-linear fashion. Motivated by theory and experiment, a power-law model $γd_0^{-2}$ is stipulated in order to be able to move forward with the assessment. This has the merit to enable us to focus on estimation of the parameter $γ$.
Thus, assuming uncorrelated input quantities, the combined standard uncertainty can be calculated from Eq. (2) using a first-order Taylor approximation.

### Velocity measurement

The velocity amplitudes of both transmitter and detector have been measured using commercial laser vibrometers with subsequent lock-in amplifiers. We identified the main contributions to the measurement uncertainty to be the linearity error (max. ± 0.1%) and calibration error (max. ± 0.5%) of the vibrometers, as specified in the data sheet of the used instruments and measurement range, the input gain accuracy of the lock-in amplifiers (max. ± 0.01%) and the angular alignment of the laser beam with respect to the detector/transmitter surface (max. ± 1°). Since the detector bending amplitude is calculated from a linear combination of three laser measurement signals, cf. section *Bending amplitude extraction*, the contribution to the combined uncertainty must be calculated accordingly.

### System parameters

The transmitter mass $m_t$ has been measured with a Mettler-Toledo XP6002S scale with reported linearity error of ±0.0005%, reproducibility of ±0.00013% and sensitivity deviation of ±0.0015%, resulting in an absolute standard uncertainty of the transmitter mass of better than 100 mg.
The distance $d_0$ was initially adjusted manually, where a systematic error of ± 0.5 mm was assumed. The automatic positioning system itself works very precisely with an error of ± 1 µm. Since $d_0$ is different for the single measurements, an average relative standard error was calculated.
The Q-factor of the detector is assumed to be constant since pressure and temperature are held constant throughout the measurements. Hence, its value is obtained from an inverse-variance weighted mean of all SDOF fit results, resulting in $Q_d = 3.587(88) \, 10^4$. The uncertainty of the resonance frequency is calculated similarly.

### SDOF Fit

The error of the measured amplitude ratio and phase shift is obtained from each SDOF fit individually. To assess the contribution to the overall combined standard uncertainty of the measured *G*, average values were calculated.

**Theoretical model**

To evaluate the relative error for estimating γ, computations of the partial derivatives of γ with respect to the beam dimensions have been performed to come up with a reasonable value. Errors due to linearization of the model or other numerical effects were not considered at this phase of the project. The standard uncertainty of γ was estimated to be 0.16% using a measurement uncertainty of 0.01 mm for all dimensions included in the model

All uncertainty contributions are summarized in the error budget presented in Extended Data Table 1. The Q-factor of the detector beam and the distance $d_0$ represent the biggest error source for the measurements presented in this paper.

**Outlook**

Due to relatively large contributions of single variables, the assessment of the uncertainty should be regarded as preliminary and yet incomplete. Therefore, measures to improve the uncertainty as well as unconsidered contributions are discussed briefly:
Using a more sophisticated beam alignment, e.g. using optical methods, the initial distance accuracy may be reduced significantly. Further, a calibration of the velocity measurement chain must be performed. According to the manufacturer, the calibration uncertainty of the laser vibrometer results almost totally from the available velocity standards. However, since we are interested in the ratio between the measured velocity signals only, a relative calibration of the measurement chains will increase the measurement accuracy, nonetheless.
Influences that are considered negligible at moment might play a role if the major uncertainty contributions decrease. This can, amongst others, comprise
- Beam parallelism
- Mass distribution of the beams
- Movement of the transmitter chamber
- Nonlinear theory of the transmitter beam displacement
- Influence of the Taylor approximation
- Model of the piezoelectric excitation

## Data availability

Source data are provided with this paper. The data that support the findings of this study are available from the corresponding author upon reasonable request.

## Acknowledgements

The authors gratefully acknowledge the support of ETH Zurich, maxon motor ag and ZC Ziegler Consultants AG.

## Author information

These authors contributed equally: Tobias Brack, Bernhard Zybach

### Affiliations

**Institute of Mechanical Systems, Department of Mechanical and Process Engineering, ETH Zurich, Tannenstrasse 3, Zürich**

Tobias Brack, Bernhard Zybach, Stephan Kaufmann Francesco Palmegiano, Jean-Claude Tomasina, Stefan Blunier, Donat Scheiwiller, Jonas Fankhauser, Jürg Dual

**Seminar for Statistics, Department of Mathematics, ETH Zurich, Rämistrasse 101, 8092 Zurich**

Fadoua Balabdaoui

### Contributions

T.B., B.Z., F.P., J.-C.T., S.B., D.S. and J.D. designed and constructed the experiment. T.B., B.Z. and J.D. conducted the experiments, T.B., F.B. and J.D. evaluated the data and analysed the results. T.B., S.K., J.F. and J.D. derived and evaluated the theory. All authors wrote and reviewed the manuscript.

### Corresponding authors

Correspondence to [Jürg Dual](#).

# Extended data figures and tables

**Extended Data Table 1:** One-sigma error budget used for the assessment of the combined measurement uncertainty of *G*.
*arithmetic mean of all measurements

|  | relative standard uncertainty (%) | Δ$G/G$ (%) |
|---|---:|---:|
| **Systematic errors** |  |  |
| Detector laser vibrometer | 0.29% | 0.21% |
| Detector laser angular misalignment | 0.01% | 0.01% |
| Detector laser lock-in amplifier | 0.01% | 0.00% |
| Transmitter laser vibrometer | 0.32% | 0.32% |
| Transmitter laser angular misalignment | 0.01% | 0.01% |
| Transmitter laser lock-in amplifier | 0.01% | 0.01% |
| Transmitter beam mass $m_t$ | 0.00% | 0.00% |
| Beam distance $d_0$ | 0.66%* | 1.31% |
| Model parameter $\gamma$ | 0.16% | 0.32% |
| **Statistical errors** |  |  |
| Fit error resonance amplitude ratio $u_{d,0}/u_{t,0}$ | 0.32%* | 0.32% |
| Detector beam resonance frequency $\omega_0$ | < 0.0001%* | < 0.0001% |
| Detector beam Q factor $Q_d$ | 2.36% | 2.36% |
| Beam distance $d_0$ | 0.0001% | 0.0002% |
|  |  |  |
|  | **combined** | **2.76%** |

**Extended Data Table 2:** Beam model parameters. Derived parameters using Euler-Bernoulli beam theory are *italicized*. Note that the transmitter resonance frequency is influenced by the driving piezo and it's backing mass.

| Parameter | Transmitter Beam | Detector Beam |
|---|---|---|
| Mass [g] | 3875.6(1) | 647.72(2) |
| Length [mm] | 1000.38(1) | 1000.00(1) |
| Width [mm] | 10.05(1) | 8.49(1) |
| Height [mm] | 20.07(1) | 16.97(1) |
| Resonance frequency @ 11.5°C [Hz] | 42.58(1) | 42.6489 ± 2E-8 |
| *Mass per unit length [kg/m]* | 3.8744(1) | 0.6477(2) |
| *Cross sectional area [$mm^2$]* | 201.7(2) | 144.1(2) |
| *Bending Stiffness [Pa $m^4$]* | 680.10(4) | 92.915(5) |
| *Young's modulus [GPa]* | 400(1) | 107.4(4) |
| *Second moment of inertia with respect to z-axis [$m^4$]* | 1.697(5) E-9 | 8.65(3) E-10 |